\documentclass[aps,showpacs,preprintnumbers,amsmath,amssymb]{revtex4}

\usepackage{graphicx}
\usepackage{epstopdf}
\usepackage{latexsym}
\usepackage{amssymb}
\usepackage{amsmath}
\usepackage{color}
\usepackage{float}
\usepackage{mathrsfs}
\usepackage[center]{subfigure}
\usepackage{dcolumn}
\usepackage[dvips]{epsfig}

\usepackage{graphicx}
\usepackage{epstopdf}
\usepackage{latexsym}
\usepackage{amssymb}
\usepackage{amsmath}
\usepackage{color}
\usepackage{mathrsfs}
\usepackage[center]{subfigure}
\usepackage{ifpdf}
\usepackage{epstopdf}

\begin{document}
\baselineskip=0.8 cm
\newcommand{\bq}{\begin{equation}}
\newcommand{\eq}{\end{equation}}
\newcommand{\bqn}{\begin{eqnarray}}
\newcommand{\eqn}{\end{eqnarray}}
\newcommand{\nb}{\nonumber}
\newcommand{\lb}{\label}
\newcommand{\PRL}{Phys. Rev. Lett.}
\newcommand{\PL}{Phys. Lett.}
\newcommand{\PR}{Phys. Rev.}
\newcommand{\PRD}{Phys. Rev. D}
\newcommand{\CQG}{Class. Quantum Grav.}
\newcommand{\JCAP}{J. Cosmol. Astropart. Phys.}
\newcommand{\JHEP}{J. High. Energy. Phys.}
\newcommand{\PLB}{Phys. Lett. B}

\title{Repulsive effect for the unbound high energy particles along the rotation axis in the Kerr-Taub-NUT spacetime}

\author{Lu Zhang$^{1,2}$, Songbai Chen$^{1,2,3}$\footnote{Corresponding author: csb3752@hunnu.edu.cn}, Jiliang
Jing$^{1,2,3}$ \footnote{jljing@hunnu.edu.cn}}

\affiliation{$^{\textit{1}}$Institute of Physics and Department of Physics, Hunan
Normal University,  Changsha, Hunan 410081, People's Republic of
China \\ $^{\textit{2}}$Key Laboratory of Low Dimensional Quantum Structures \\
and Quantum Control of Ministry of Education, Hunan Normal
University, Changsha, Hunan 410081, People's Republic of China\\
$^{\textit{3}}$Synergetic Innovation Center for Quantum Effects and Applications,
Hunan Normal University, Changsha, Hunan 410081, People's Republic
of China}

\begin{abstract}
\baselineskip=0.6 cm
\begin{center}
\end{center}

We have investigated the acceleration of the unbound high energy particles moving along the rotation axis in the Kerr-Taub-NUT spacetime, and then study the dependence of the repulsive effects on the NUT charge for the particles in the spacetime. Whether the repulsive effects with the NUT charge become stronger depends on the Carter constant, the position and velocity of the particles themselves.
We also present numerically the changes of the observable velocity and acceleration with the NUT charge for the unbound particles in the Kerr-Taub-NUT spacetime.
\end{abstract}

\maketitle

\newpage
\section{Introduction}

Black hole jets are one of the most spectacular astronomical sights in the sky, which are typically associated with outflows of matter emitting as an extended beam along the rotation axis of black hole. These jets are relativistic since the matter in the beam always can be accelerated to nearly the speed of light. The acceleration process of particles in the jets is so complicated that it is not fully understand at present. A lot of current theoretical models accounted for such kind of jets are based on the electromagnetic field whose magnetic field lines anchored in the  horizon of black hole or in the accretion disk \cite{Blandford1,Blandford2,Punsly} and then the matter in the beam are accelerated by extracting the corresponding energy. The Blandford-Znajek process \cite{Blandford1} is such a mechanism for the acceleration process of particles by extracting the energy from magnetic fields around an accretion disk, which are dragged and twisted by the spin of the black hole. Penrose process is another important acceleration mechanism for the particles in jets in which the energy is extracted from a rotating black hole by frame dragging \cite{Penrose1,Penrose2,Penrose3}. With Penrose process, the participation of neutral particles are allowable in the formation of astrophysical jets since Penrose process can occur without electromagnetic fields. However,
it is still lack of a full mechanism to explain the formation and powering of the black hole jets. 

It is worthwhile looking for other types of mechanisms since the astrophysical jets may be caused by the comprehensive effect of various physical processes near a black hole. B. Mashhoon \textit{et al} \cite{BM1,BM2,BM12,BM13,BM3,BM32} studied the kinematical effects associated with the congruence of particle world lines and found that the behavior of the components of the deviation vector connecting any two neighboring world lines during the evolution play important role in the formation of the collimated jet. For example, 
the gravitational tidal effect by a central rotating object can change the acceleration of a beam of ultrarelativistic particles and modify the dynamics of relativistic flows near black holes \cite{BM2,BM12,BM13}. For an initially spherical bunch of particles along the axis of a rotating source, the inertial strains and curvature complete to generate jet-like structures \cite{BM3}. These results are very useful for us to understand more about the formation and powering of astrophysical jets.

Recently, the repulsive effect near axis of black hole has been proposed to explain the formation of a highly energetic and collimated jet powered by a rotating black hole \cite{Gariel1,Gariel2,Gariel3,Anz1,Anz2,Herrera, Opher}. In this purely gravitational theoretical model, a jet is represented as a set of test particles following the unbound geodesics near the rotation axis of a black hole. After undergoing Penrose process, these high energies particles are ejected along the collimated geodesics due to a powerful gravitational effect of repulsion near the axis because of their outward acceleration. Considered that such an outward acceleration could be very big, with this purely gravitational model, it is possible in principle to obtain a collimated and energetic
jet with some generic features.  Especially,
the presence of a characteristic radius and of the size of
the ergosphere around the most energetic
particles, might be observationally testable \cite{Gariel1}.
Although it focuses only on the mechanism of ejection of a single particle of the bunch and requires Penrose
process to produce a suitable and rather special distribution of
outgoing particles, this theoretical model of repulsive effect has been applied widely in the Kerr black hole spacetime \cite{Gariel1,Gariel2,Gariel3,Anz1,Anz2,Gariel4} including the analysis of the profile of the M87 jet \cite{Gariel4} and the study of the observable velocities and accelerations for the particles near the rotation axis of black hole \cite{Anz2}. As in other purely gravitational theoretical models, it does not resort to magnetic field, which ensures the participation of neutral particles in the astrophysical jets.

In general relativity, Kerr-Taub-NUT (KTN) spacetime is another important stationary and axisymmetric  solution of Einstein field equations for electro-vacuum spacetime possessing with gravitomagnetic monopole and dipole moments \cite{KTN1,KTN2}. Comparing with the usual Kerr case, the KTN spacetime carries with an extra NUT charge, $n$, which  plays the role of a magnetic mass inducing a topology in the Euclidean section. The presence of the
NUT charge results in that the spacetime structure becomes asymptotically non-flat and  there are conical singularities on its axis of symmetry. Although there exist these undesired features, the KTN spacetime still is attractive for exploring various physical phenomena in general relativity \cite{KTNapp1,KTNapp2,KTNapp3,KTNapp4,KTNapp5,KTNapp6,KTNapp7,KTNapp8,KTNapp9}.
 As in the previous discussion, the formation of astrophysical jets involves magnetohydrodynamics, relativistic gravity and the kinematical effects associated with the congruence of particle world lines including the relative acceleration effects and tidal forces.
In this paper, we will focus only on the purely gravitational acceleration model analyzed in Refs. \cite{Gariel1,Gariel2,Gariel3,Anz1,Anz2,Herrera, Opher} and study the acceleration of the unbound high energy particles moving along the rotation axis in the KTN spacetime, and then to see what effects of the NUT charge on the repulsive effects for the particles near the rotation axis of black hole.

This paper is organized as follows. In Sec. II we derive briefly the timelike geodesics in the KTN spacetime with  the Weyl coordinates and then study the acceleration of the timelike particles along the rotation axis. In Sec. III, we study the dependence of the repulsive effect on the NUT charge for the unbound high energy particles in the KTN spacetime. In Sec. IV, we present numerically the effects of the NUT charge on the observable velocity and acceleration for the unbound high energy particles. Finally, we present a brief summary.

\section{Acceleration of the timelike particles along the rotation axis in the Kerr-Taub-Nut spacetime}

In the Boyer-Lindquist coordinates, the general KTN solution of the Einstein field equations is described by the metric \cite{KTN1,KTN2}
\begin{eqnarray}
ds^2=&-&\frac{{\bar r}^2-2M{\bar r}+a^2-n^2}{{\bar r}^2+(n+a\cos\theta)^2}\bigg[dt
+(2n\cos\theta-a\sin^2\theta )d\phi\bigg]^2
+\frac{\sin^2\theta}{{\bar r}^2+(n+a\cos\theta)^2}\times
\nonumber\\&&\bigg[a\, dt-({\bar r}^2+a^2+n^2)\,d\phi\bigg]^2 +\bigg[{\bar r}^2+(n+a\cos\theta)^2\bigg](\frac{d{\bar r}^2}{{\bar r}^2-2M{\bar r}+a^2-n^2}+d\theta^2),
\label{1}
\end{eqnarray}
where $M$, $a$, and $n$ are the mass, the rotation parameter, and the NUT charge of the source, respectively.
The positions of event horizon and Cauchy horizon of the KTN spacetime are located at
\begin{eqnarray}
\bar{r}_{H,C}=M\pm\sqrt{M^2-a^2+n^2},
\end{eqnarray}
 which are the roots of the equation
\begin{eqnarray}
{\bar r}^2-2M{\bar r}+a^2-n^2=0.
\end{eqnarray}
With increase of the NUT parameter $n$, it is obvious that the radius of the event
horizon becomes larger.

The geodesic equation of a timelike particle in the curve spacetime is
\begin{eqnarray}
\frac{d^2x^{\mu}}{d\lambda^2}+\Gamma^{\mu}_{\alpha\beta}
\frac{dx^{\alpha}}{d\lambda}\frac{dx^{\beta}}{d\lambda}=0,
\end{eqnarray}
with
\begin{eqnarray}
g_{\mu\nu}\frac{dx^{\mu}}{d\nu}\frac{dx^{\beta}}{d\lambda}=-1.
\end{eqnarray}
 Since the metric coefficients in the KTN spacetime (\ref{1}) are independent of the coordinates $t$ and $\phi$,  there exist
the conserved specific energy $E$ and the conserved $z$-component
of the specific angular momentum $L$ at infinity  for the geodesic motion of particle. With these two conserved quantities and rescaling the $\bar r$ coordinate as $r={\bar r}/M$ \cite{Anz1,Anz2}, one can find that the polar and angular geodesic equations for a timelike particle can be expressed as
\begin{eqnarray}
{\dot r}^2=\frac{M^2[a_4r^4+a_3r^3+a_2r^2+a_1r+a_0]}{
[M^2r^2+(n+a\cos\theta)^2]^{2}},
\label{2}
\end{eqnarray}
and
\begin{eqnarray}
{\dot\theta}^2= \frac{M^2[b_4\cos^4\theta+b_3\cos^3\theta+b_2\cos^2\theta
+b_1\cos\theta+b_0]}{(1-\cos^2\theta)[M^2r^2+(n+a\cos\theta)^2]^{2}}, \label{3}
\end{eqnarray}
respectively.
Here the dot stands for derivatives with respect to the proper time $\tau$.  The coefficients $a_i$ and $b_j$ are given by
\begin{eqnarray}
&&a_0=\frac{n^2 \left(3 a^2 E^2-a^2-4 a E L+L^2+Q\right)-a^2 Q+\left(E^2+1\right) n^4}{M^{4}}, \label{4}\\
&&a_1=\frac{2 \left[(a E-L)^2+n^2+Q\right]}{M^2}, \label{5}\\
&&a_2=\frac{a^2 \left(E^2-1\right)-L^2+2 E^2 n^2-Q}{M^2}, \label{6}\\
&&a_3=2, \label{7}\\
&&a_4=E^2-1, \label{8}
\end{eqnarray}
and
\begin{eqnarray}
&&b_0=\frac{Q}{M^2},\;\;\;\;\; \label{9}\\
&&b_1=\frac{2 n \left(2 a E^2-a+2 E L\right)}{M^2}, \label{09}\\
&&b_2=\frac{a^2 \left(E^2-1\right)-L^2-4 E^2 n^2-Q}{M^2}, \label{10}\\
&&b_3=-\frac{2 a \left(2 E^2-1\right) n}{M^2}, \label{010}\\
&&b_4=-\left(\frac{a}{M}\right)^2\left(E^2-1\right) , \label{11}
\end{eqnarray}
where $Q$ is the Carter constant. As the NUT charge $n$ tends to zero, these coefficients and the corresponding geodesic equations reduce to that in the usual Kerr black hole case \cite{Anz1,Anz2}. The presence of the NUT charge $n$ yields that the coefficient $a_2$ is no longer equal to $b_2$, which means that the NUT charge should bring some new effects on the acceleration of the timelike particles moving along the symmetry axis. The other two-component equations of the timelike geodesics can be written as
\begin{eqnarray}
 \tilde{\rho}^{2}\dot{t}=&-&E\bigg\{\sin^{2}\theta(a+2n\csc^{2}\theta)^{2}
 -4n[n+a(1+\cos\theta)]\bigg\}
                   -(2n \cos\theta \csc^{2}\theta-a)L\nonumber\\
                 &+&\frac{(r^{2}M^{2}+a^{2}+n^{2})[E(r^{2}M^{2}+a^{2}+n^{2})-aL]}{\Delta},\\
 \tilde{\rho}^{2}\dot{\phi}=&-&(a-2n\cot\theta\csc\theta)E+
 L\csc^{2}\theta+\frac{a[E(r^{2}M^{2}+a^{2}+n^{2})-aL]}{\Delta},
  \label{Phidot}
\end{eqnarray}
with
\begin{eqnarray}
 \tilde{\rho}^{2}&=&r^{2}M^{2}+(n+a\cos\theta)^{2},\\
 \Delta&=&r^{2}M^{2}-2M^{2}r+a^{2}-n^{2}.
\end{eqnarray}
Similarly, one can find that the NUT charge also changes these two geodesic equations. As in the Kerr black hole case \cite{Anz1,Anz2}, we here focus only on the particles on unbound geodesics with $E\geq 1$ \cite{Chandra} since the particles in the jets own very high energies. For an  axially symmetric spacetime, it is useful to adopt the Weyl cylindrical coordinates $(\rho, z, \phi)$ to investigate the motions of particles moving along the rotation axis \cite{Gariel1}. In the KTN spacetime, the dimensionless Weyl cylindrical coordinates in multiples of geometrical units of mass $M$ can be expressed as
\begin{equation}
\rho=\left[(r-1)^2-A\right]^{1/2}\sin\theta, \;\; z=(r-1)\cos\theta, \label{12}
\end{equation}
with
\begin{equation}
A=1-\frac{a^2-n^2}{M^2}. \label{13}
\end{equation}
Defining the quantity
\begin{equation}
\alpha=\frac{1}{2}\left\{\left[\rho^2+(z+\sqrt{A})^2\right]^{1/2}+
\left[\rho^2+(z-\sqrt{A})^2\right]^{1/2}\right\}, \label{16}
\end{equation}
one can obtain the inverse transformation of the Eq.(\ref{12}),
\begin{eqnarray}
\sin\theta &=&\frac{\rho}{(\alpha^2-A)^{1/2}}, \;\; \;\; \;\; \;\; \;\; \;\; \;\; \;\; \;\;\cos\theta=\frac{z}{\alpha},\label{15}\\
r&=&\alpha +1,\;\; \;\; \;\; \;\; \;\; \;\; \;\; \;\; \;\;z=\left(1-\frac{\rho^2}{\alpha^2-A}\right)^{1/2}\alpha. \label{15a}
\end{eqnarray}
For KTN spacetime, the condition $A\geq 0$ ensures disappearance of the naked singularity. As $\alpha$ and $r$ are constants, one can find that the curves obeyed Eq.(\ref{16}) are ellipses with the semi-major axis $\alpha$ and the eccentricity $e={\sqrt A}/\alpha$ in the $(\rho,z)$ plane, which means that these curves approach  to circles in the cases with the large value of $\alpha$. With the methods adopted in Refs.\cite{Anz1,Anz2}, one can obtain the the geodesic equations of the timelike particles in the Weyl cylindrical coordinates, which take the forms
\begin{eqnarray}
 M{\dot\rho}&=&\frac{1}{U}\left[\frac{P\alpha^3\rho}{\alpha^2-A}
+\frac{S(\alpha^2-A)z}{\alpha\rho}\right], \label{dotrho}\\
 M{\dot z}&=&\frac{1}{U}\,(Pz-S)\alpha, \label{dotz}\\
M\dot{\phi}&=&-\frac{\alpha^2E}{MU}\left[a-\frac{2nz\left(\alpha^2-A\right)
}{a\rho^2}\right]
 +\frac{\alpha^2L\left(\alpha^2-A\right)}{MU\rho^2}+\frac{\alpha^2a
 \left\{E\left[\left(\alpha+1\right)^{2}M^{2}
 +a^{2}+n^{2}\right]-aL\right\}}{MU\Delta},\label{Phidot1}\\
\dot{t}=&-&\frac{E\alpha^2}{M^2U}\left\{\frac{\rho^2}{\alpha^2-A}
(a+2n\frac{\alpha^2-A}{\rho^2})^{2}-4n\left[n+a(1+\frac{z}{a})\right]\right\}
-\frac{L\alpha^2}{M^2U}\left[\frac{2nz\left(\alpha^2-A\right)}{\alpha\rho}-a\right]\nonumber\\
&+&\frac{\alpha^2\left[\left(\alpha+1\right)^{2}M^{2}+a^{2}+n^{2}\right]\left\{E\left[\left(\alpha+1\right)^{2}M^{2}
+a^{2}+n^{2}\right]-aL\right\}}{M^2U\Delta},\label{tdot}
\end{eqnarray}
where
\begin{eqnarray}
&&P=\left[a_4(\alpha+1)^4+a_3(\alpha+1)^3+a_2(\alpha+1)^2 +a_1(\alpha+1)+a_0\right]^{1/2}, \label{19}\\
&&S=-\left(b_4z^4+b_3\alpha z^3+b_2\alpha^2z^2+b_1\alpha^3z+b_0\alpha^4\right)^{1/2}, \label{20}\\
&&U=(\alpha+1)^2\alpha^2+\frac{\left(n\alpha+az\right)^2}{M^2}, \label{21}\\
&&\Delta=\left(\alpha+1\right)^{2}M^{2}-2M^{2}\left(\alpha+1\right)+a^{2}-n^{2}.\label{Delta2}
\end{eqnarray}
Here the signs of $P$ and $S$ are chosen to indicate outgoing particle
geodesics. Making use of Eqs. (\ref{16}), (\ref{dotrho}) and (\ref{dotz}), one has
\begin{equation}
M{\dot\alpha}=\frac{P\alpha^2}{U}. \label{22}
\end{equation}
Taking the second proper time derivative of $z$ from Eq.(\ref{dotz}), one can obtain the acceleration of a test massive particle along the $z$ axis in the KTN spacetime
\begin{eqnarray}
MU^2{\ddot z}&=&
\left[UP-\frac{2a(Pz-S)(n\alpha+az)}{M^2}\right]\alpha{\dot z}
+(Pz-S)\left[U-2\alpha^2(\alpha+1)(2\alpha +1)-\frac{2n\alpha(n\alpha+az)}{M^2}\right]{\dot \alpha}\nonumber\\&&
+U({\dot P}z-{\dot S})\alpha. \label{23}
\end{eqnarray}
From Eqs. (\ref{dotz}), (\ref{19}), (\ref{20}), and (\ref{22}),  we have
\begin{eqnarray}
MU({\dot P}z-{\dot S})&=&\frac{1}{2}\left[4a_4(\alpha +1)^3+3a_3(\alpha +1)^2
+2a_2(\alpha +1)+a_1\right]\alpha^2z
-2PS\alpha\nonumber\\&&+\frac{\alpha}{2}(4b_4z^3+3b_3z^2\alpha+2b_2z\alpha^2+b_1\alpha^3). \label{24}
\end{eqnarray}
Substituting (\ref{dotz}), (\ref{22}) and (\ref{24}) into (\ref{23}),  we find that Eq. (\ref{23}) can be simplified further as
\begin{eqnarray}
\frac{M^2U^3}{\alpha^2}\;{\ddot z}
&=&-P S \left[2 (\alpha +1) \alpha ^2+\frac{2 \alpha  n \left(a z+\alpha  n\right)}{M^2}\right]-\frac{S^2 \left[2 a \left(a z+\alpha  n\right)\right]}{M^2}-2 (\alpha +1) \alpha ^3 P^2 z\nonumber\\&&+\frac{1}{2} \alpha  U z \left[4 a_4 (\alpha +1)^3+3 a_3 (\alpha +1)^2+2 a_2 (\alpha +1)+a_1\right]\nonumber\\&&+\frac{1}{2} U \left[\alpha ^3 b_1+4 b_4 z^3+3 \alpha  b_3 z^2+2 \alpha ^2 b_2 z\right]. \label{28}
\end{eqnarray}
Solving this differential equation, we can study the acceleration of a test massive particle along the $z$ axis in the KTN spacetime. Clearly,  the presence of the NUT charge will change the accelerations of the particles. In the next section, we will study the repulsive effect for the timelike particles and to see how the NUT charge affect this effect in the KTN spacetime.

\section{Repulsive effect for the unbound high energy particles along the rotation axis in the Kerr-Taub-NUT spacetime}

The existence of the ergosphere is an important feature of a rotating axial symmetrical spacetime. Thus, we first consider the case in which the particle lies in the ergosphere surface as in the Kerr case \cite{Anz1}. In the Boyer-Lindquist coordinates, the outer surface of the ergosphere for the metric (\ref{1}) is given by
\begin{equation}
r=1+\left(1-\frac{a^2\cos^2\theta-n^2}{M^2}\right)^{1/2}. \label{32a}
\end{equation}
Transforming it into the Weyl coordinates (\ref{15})-(\ref{15a}), one can find that it becomes
\begin{equation}
z^2=\left[1-\rho_e^2\left(1-\frac{\rho}{\rho_e}\right)
+\frac{n^2}{M^2}\right]\left(1-\frac{\rho}{\rho_e}\right), \label{33a}
\end{equation}
with $\rho_e\equiv a/M$, which is the value of $\rho$ for which the ergosphere
cuts the equatorial plane $z=0$. Thus, for ergosphere surface, the value of $\rho$ is in the range $(0, \;\rho_e)$. Let us now to discuss the acceleration $\ddot{z}$ ( along the $z$ axis ) for the timelike particles leaving the ergosphere in these two limit cases $\rho=0$ and $\rho=\rho_e$.
For the case $\rho=0$, from Eqs.(\ref{16}), (\ref{20}), and (\ref{33a}), it is easy to obtain
\begin{equation}
\alpha=z=\sqrt{A}, \label{34a}
\end{equation}
and
\begin{equation}
S=-(b_4+b_3+b_2+b_1+b_0)^{1/2}A=-\frac{A\sqrt{-(L-2nE)^2}}{M}, \label{35a}
\end{equation}
It implies that only the particles with $L=2nE$ can leave the ergosphere along geodesics at $\rho=0$, which is different from that in the case of a usual Kerr black hole in which the condition of particle leaving the ergosphere at $\rho=0$ is $L=0$ \cite{Anz1}. This means that the NUT charge $n$ will change acceleration of a timelike particle and modify further the repulsive effect. From Eqs. (\ref{28}) and (\ref{34a}) we have
\begin{eqnarray}
\ddot{z}&=&\frac{1}{2W^3}\bigg[W_0+W_1\sqrt{(1-A)M^2+n^2}
-\sqrt{A}QW\bigg],\label{36a}
\end{eqnarray}
where
\begin{eqnarray}
W&=&(1+\sqrt{A}) M^2+n^2+n\sqrt{(1-A)M^2+n^2},\nonumber\\
W_0&=&(1+\sqrt{A})^2\sqrt{A}
   M^4[(E^2-1)\sqrt{A}-E^2]+(1+\sqrt{A}) M^2 [\sqrt{A}(1-6E^2)-4E^2\nonumber\\&+&A(5E^2-2)]n^2 +[\sqrt{A}-(4+9\sqrt{A})E^2]n^4,\nonumber\\
W_1&=&(1+\sqrt{A})M^2[4E^2+A(E^2-1)+
    \sqrt{A}(1+3E^2)]n+[(4-\sqrt{A})E^2+\sqrt{A}]n^3.
\end{eqnarray}
From Eq. (\ref{36a}), it is easy to find that when the Carter constant satisfies
\begin{eqnarray}
Q<Q_c\equiv\frac{W_0+W_1\sqrt{(1-A)M^2+n^2}}{\sqrt{A}W},
\end{eqnarray}
the acceleration along $z$ direction $\ddot{z}$ for the particles on the ergosphere surface at $\rho=0$ is always positive. As the NUT charge $n$ vanishes, the critical value $Q_c$ becomes
\begin{eqnarray}
Q_c=-M^2[1+\sqrt{A}+\rho^2_e(E^2-1)],
\end{eqnarray}
which is consistent with that in the background of a Kerr black hole \cite{Anz1}. Considering a physical fact that the NUT charge is small,  one can obtain
\begin{eqnarray}
Q_c=-M^2[1+\sqrt{A}+\rho^2_e(E^2-1)]+\frac{(\sqrt{A}+1)
[4E^2(\sqrt{A}+1)+\sqrt{A}]Mn}{\sqrt{A}} +\mathcal{O}(n^2),
\end{eqnarray}
which means that the small $n$ extends the range of $Q$ in which $\ddot{z}>0$ and then the more particles can be ejected along the $z-$direction in the KTN spacetime. Moreover, in this small $n$ case, we have
\begin{eqnarray}\label{45}
\ddot{z}&=&-\frac{\sqrt{A}}{2(1+\sqrt{A})^2M^2}
\bigg[1+\sqrt{A}+\rho^2_e(E^2-1)+\frac{Q}{M^2}\bigg]+
\frac{\sqrt{1-A}}{2(1+\sqrt{A})^3M^5}\times
\nonumber\\&&\bigg(2\sqrt{A}Q+(1+\sqrt{A})M^2[4E^2(1-A)+
   2A(E^2+1)+\sqrt{A}(6E^2+1)]\bigg)
n+\mathcal{O}(n^2).
\end{eqnarray}
Since the coefficient of $n$ in right-hand-side of Eq.(\ref{45}) always is positive for the unbound particles $E\geq1$, the presence of the NUT charge $n$ enhances the acceleration $\ddot{z}$ and make the repulsive effect more stronger in this case.

For the particle leaving the ergosphere at $\rho=\rho_e$ (\ref{33a}), one can obtain
\begin{equation}
z=0, \;\;\;\;\;\;\;\; \alpha=\sqrt{1+\frac{n^2}{M^2}}, \label{38a}
\end{equation}
and
\begin{equation}
S=-\left(1+\frac{n^2}{M^2}\right)\frac{\sqrt Q}{M}, \label{39a}
\end{equation}
which implies the condition of particles leaving the ergosphere at $\rho=\rho_e$ is still
$Q>0$ and the presence of $n$ does not change this condition.
From Eq.(\ref{28}), we obtain
\begin{eqnarray}
\ddot{z}&=&\frac{1}{4\sqrt{M^2+n^2}(M+\sqrt{M^2+n^2})^3}
\bigg[(M^2+M\sqrt{M^2+n^2}+2n^2)\sqrt{W_2Q}
\nonumber\\&+&n\bigg(\frac{(M+\sqrt{M^2+n^2})
[2LE+M\rho_e(2E^2-1)]}{M}-\frac{\rho_eQ}{\sqrt{M^2+n^2}}\bigg)
\bigg],
\end{eqnarray}
with
\begin{eqnarray}
W_2&=&2\sqrt{M^2+n^2}(M+\sqrt{M^2+n^2})\bigg[
   2 M(\sqrt{M^2+n^2}E-L\rho_e )E+2(M^2+n^2)E^2\nonumber\\&+&
   M^2\rho^2_e(2E^2-1)\bigg]-\rho^2_eM^2Q.
\end{eqnarray}
In the small $n$ case, we have
\begin{eqnarray}\label{49}
\ddot{z}&=&\frac{\sqrt{Q}}{M^3}\bigg[16E^2+4\rho^2_e(2E^2-1)
-\frac{8EL\rho_e}{M}-\frac{\rho^2_eQ}{M^2}\bigg]^{1/2}
+\bigg[\frac{4ELM+2M^2\rho_e(2E^2-1)-\rho_eQ }{32M^5}\bigg]n\nonumber\\&&+\mathcal{O}(n^2).
\end{eqnarray}
In order to ensure the quantity in the square root in Eq.(\ref{49}) is positive, the Carter constant $Q$ must satisfies
\begin{eqnarray}
 Q<Q_{c2}=4M \bigg[M(2E^2-1)-\frac{2EL}{\rho_e}+\frac{4ME}{\rho^2_e}\bigg].
 \end{eqnarray}
Moreover,  the condition of the positive coefficient of $n$ in Eq.(\ref{49})
is
\begin{eqnarray}
 Q<Q_{c3}=2M \bigg[M(2E^2-1)+\frac{2EL}{\rho_e}\bigg].
\end{eqnarray}
This means that in the small $n$ case with the increase of the NUT charge, the acceleration $\ddot{z}$ increases and the repulsive effect becomes stronger as $Q<Q_{c3}<Q_{c2}$. Similarly, as the Carter constant $Q_{c3}<Q<Q_{c2}$, the acceleration $\ddot{z}$ decreases and the corresponding repulsive effect becomes weaker.

Finally, let us to consider the particles moving near the equatorial plane as in the Kerr case \cite{Anz1}. In this limit, we have $z=0$. This means that the acceleration of a test particle along $z$ direction can be written as
\begin{equation}
\ddot z=-\frac{2 \left(a \alpha ^3nS^2+\alpha ^4 n^2 P S\right)}{M^4 U^3}+\frac{\alpha ^5 b_1}{2 M^2 U^2}-\frac{2 (\alpha +1) \alpha ^4 P S}{M^2 U^3}, \label{s1}
\end{equation}
and its velocity along $\rho$- and $z$- directions become
\begin{equation}
\dot\rho=\frac{P\alpha^3\rho}{MU(\alpha^2-A)}, \;\; \;\; \;\; \;\;\;\; \;\; \;\; \;\; \dot z=-\frac{S\alpha}{MU}. \label{s2}
\end{equation}
Substituting (\ref{s2}) into (\ref{s1}), we obtain
\begin{eqnarray}
\ddot{z}&=&\frac{1}{\alpha^2\left[(\alpha +1)^2 M^2+n^2\right]^2} \bigg\{2\rho\dot{\rho}\dot{z}[M^2(1+\alpha)+n^2] [M^2(1+\alpha)^2+n^2]\nonumber \\&+&\alpha n\bigg[2EL\alpha^2-2a\dot{z}^2[M^2(1+\alpha)^2+n^2]+a\alpha^2(2E^2-1)\bigg]
\bigg\}.\label{s3}
\end{eqnarray}
Comparing with that in the Kerr black hole case \cite{Anz1}, it becomes more complicated. In the small $n$ case, it is dominated by
\begin{eqnarray}
\ddot{z}&=&\frac{2\rho\dot{\rho}\dot{z}}{\alpha^2(\alpha+1)}
+\bigg[\frac{2EL\alpha^2-2a\dot{z}^2M^2(1+\alpha)^2+
a\alpha^2(2E^2-1)}{M^4\alpha(1+\alpha)^4}\bigg]n+\mathcal{O}(n^2).
\end{eqnarray}
Obviously, the NUT charge enhances the acceleration $\ddot{z}$ as $\dot{z}^2<\frac{\alpha^2 [2EL+a(2E^2-1)]}{2aM^2 (1+\alpha)^2}$ and suppresses the acceleration $\ddot{z}$ as $\dot{z}^2>\frac{\alpha^2 [2EL+a(2E^2-1)]}{2aM^2 (1+\alpha)^2}$. Thus, for the particles moving near the equatorial plane, the effect of the NUT charge on the acceleration $\ddot{z}$ and the corresponding repulsive effect depends on their $z$-component of velocities $\dot{z}$.

\section{Observable Velocity and Acceleration for the unbound high energy particles in  the Kerr-Taub-NUT spacetime}
In this section, we will consider the observable velocity and acceleration along $z$ axis for the particles in the Kerr-Taub-NUT spacetime.
As in Ref.\cite{Anz2},
the dimensionless observable velocity and acceleration can be, respectively, defined as
\begin{equation}
\beta_z\equiv\frac{v_{z}}{c}=\frac{dz}{dt}=\frac{\dot z}{\dot t}, \label{vz}
\end{equation}
and
\begin{equation}
\beta_z^{'}\equiv\frac{d\beta_z}{dt}=\frac{1}{\dot t}\frac{d}{d\tau}\left(\frac{\dot z}{\dot t}\right). \label{az}
\end{equation}
In order to study the effects of the NUT charge $n$ on these observables, we must solve the geodesic equations (\ref{dotrho})-(\ref{tdot}).
From Eqs.(\ref{dotrho}) and (\ref{dotz}), we have
\begin{equation}
\frac{d\rho}{dz}=\frac{P\alpha^4\rho^2+S(\alpha^2-A)^2z} {%
(Pz-S)(\alpha^2-A)\alpha^2\rho}.  \label{Traject}
\end{equation}
Here we restrict our study to the quadrant $\rho>0$ and $z>0$ in
the projected meridional plane.
Solving this differential equation with some initial conditions, one can get the trajectories of the particles in the spacetime. Here, we focus on the special geodesics accounted for a perfect collimation of jets as in Ref.\cite{Anz2}. As $z\rightarrow\infty$, the quantity $\rho $ for these special geodesics tends to
\begin{eqnarray}
\rho_1\equiv\lim_{z\rightarrow\infty}\rho=\left(\frac{4b_0+3b_1+2b_2+
b_3}{2a_4}\right)^{1/2}=\rho_e\left[1+\frac{Q+4n^2E^2+
2an\left(2E^2-1\right)}{a^2\left(E^2-1\right)}\right]^{1/2},  \label{Rho1}
\end{eqnarray}
where $L$ has been taken to be $L=2nE$ in order to ensure that $S$ is real in the KTN spacetime. Actually, the limit value $\rho_1$ is a characteristic parameter of the collimated jet's ejection as discussed in Ref.\cite{Gariel1}. It is obvious that
the characteristic parameter $\rho_1$ increases with the NUT charge $n$ for fixed $M$, $a$, $Q$, and $E$.
\begin{figure}
\centering
\includegraphics[width=6cm]{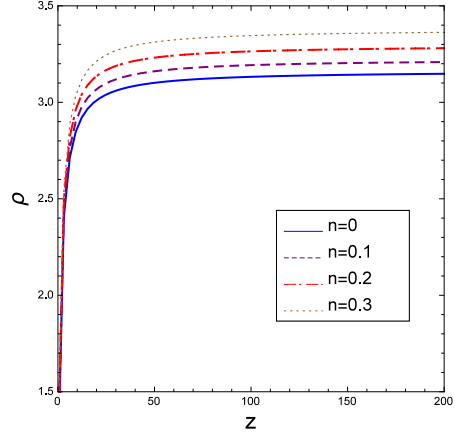}
\caption{The changes of $\rho$ with $z$ for the different $n$ and for the fixed parameters $a=0.877004$, $E=10^{6}$, and $Q=9.238849$. }
\label{Fig1}
\end{figure}

\begin{figure}
\centering
\includegraphics[width=6cm]{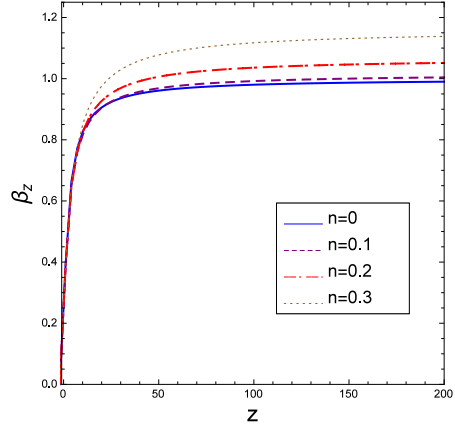}
\caption{The observable velocity $\beta_z$ in the $z-$axial direction for the timelike particle moving along the geodesics shown in Fig.(\ref{Fig1}).}
\label{Fig2}
\end{figure}

\begin{figure}
\centering
\includegraphics[width=6cm]{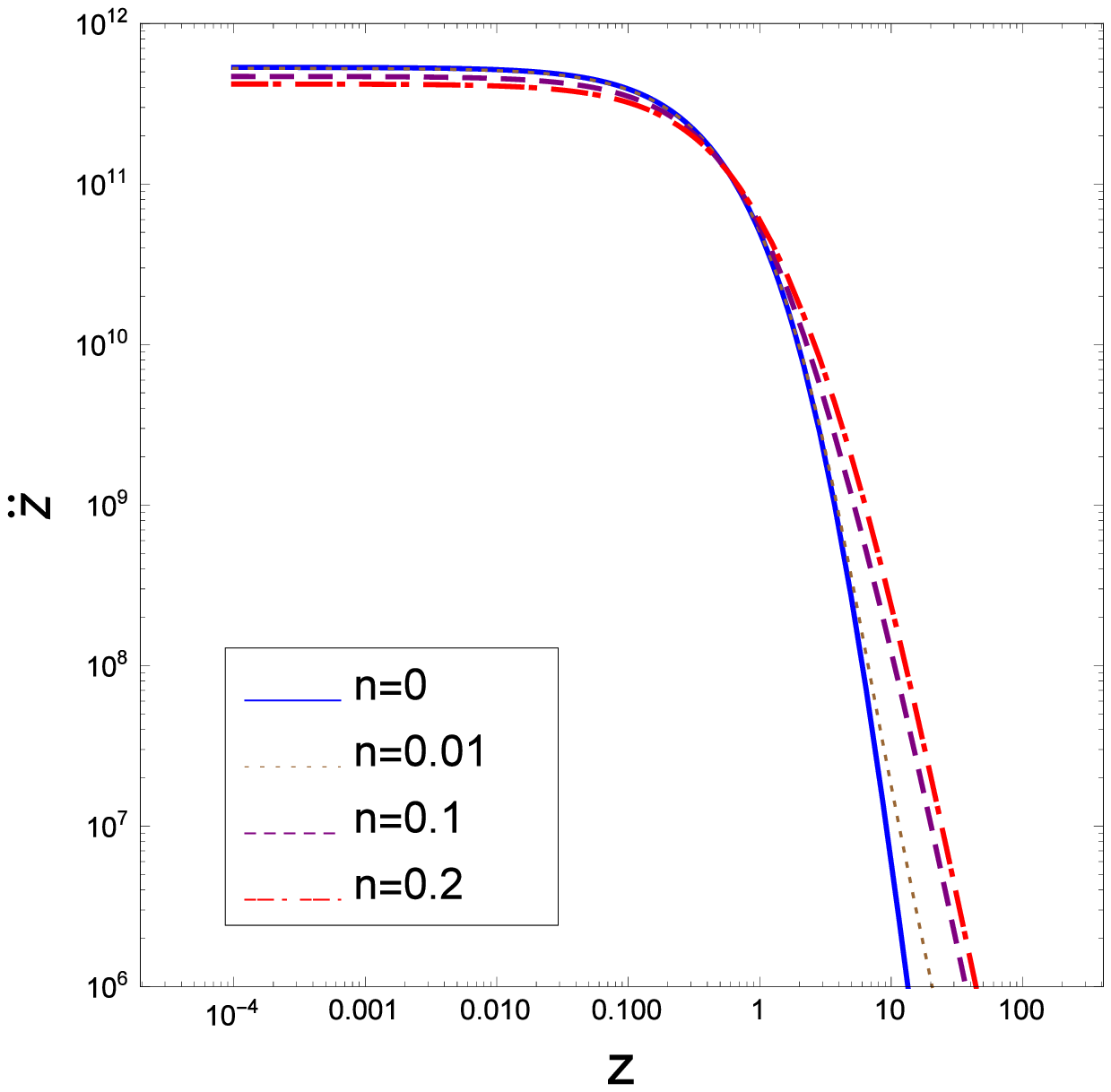}\includegraphics[width=6.5cm]{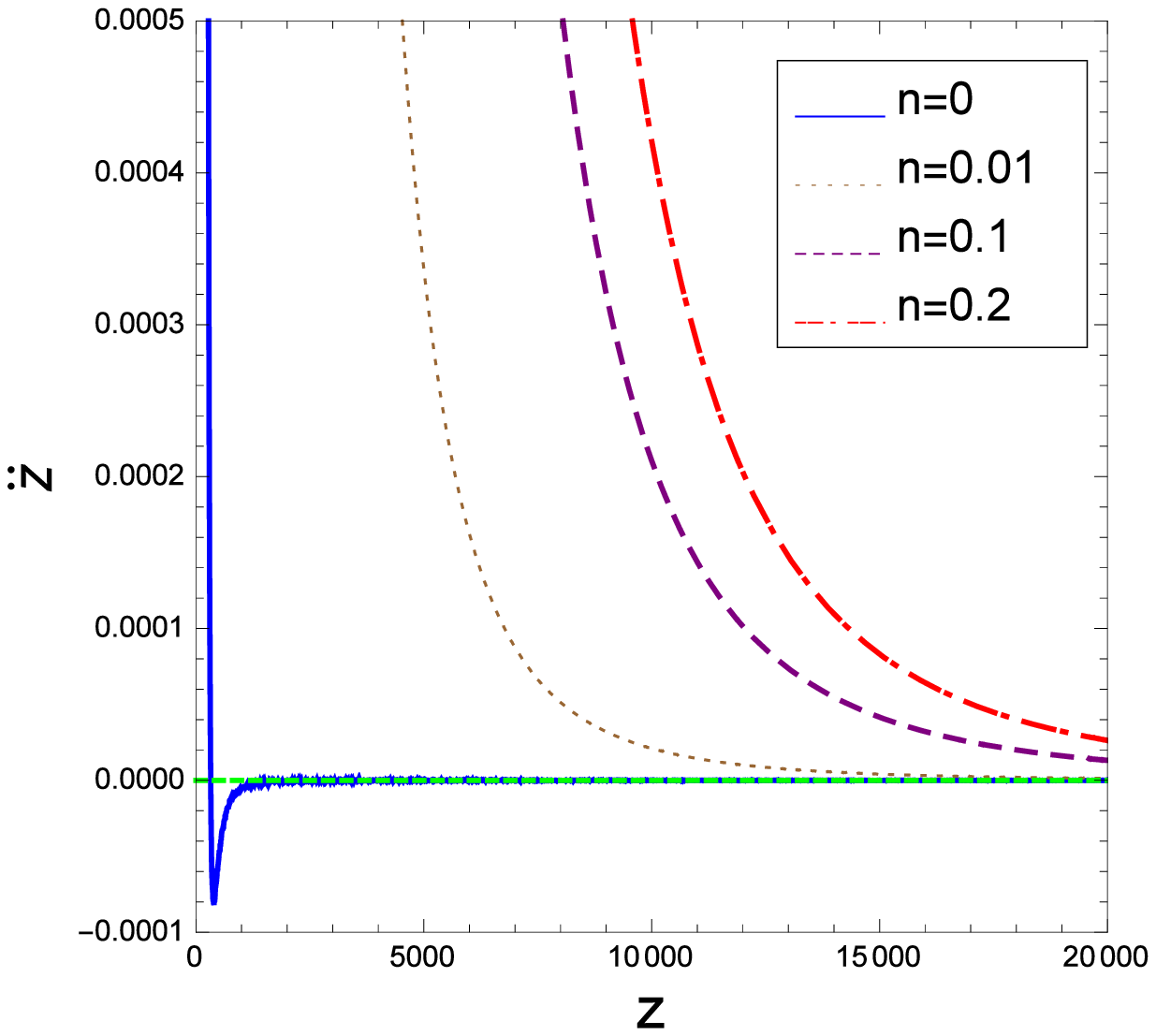}
\caption{The change of the proper acceleration along axis $\ddot{z}$ with $z$ for different $n$. The figure is  plotted with LogLog scale in the left panel and with linear scale in the right one. }
\label{Fig3}
\end{figure}

\begin{figure}
\centering
\includegraphics[width=6cm]{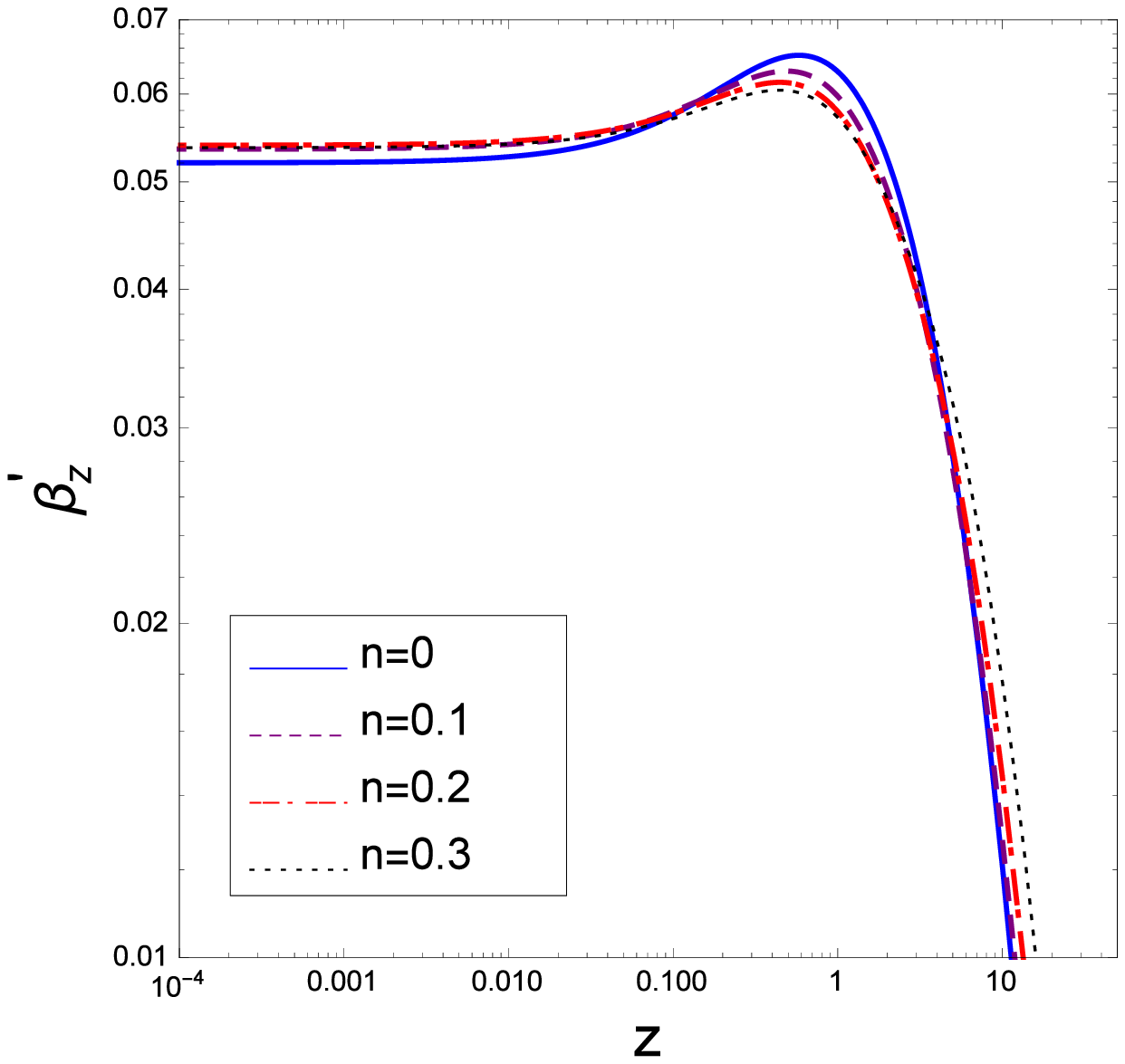}\includegraphics[width=6.1cm]{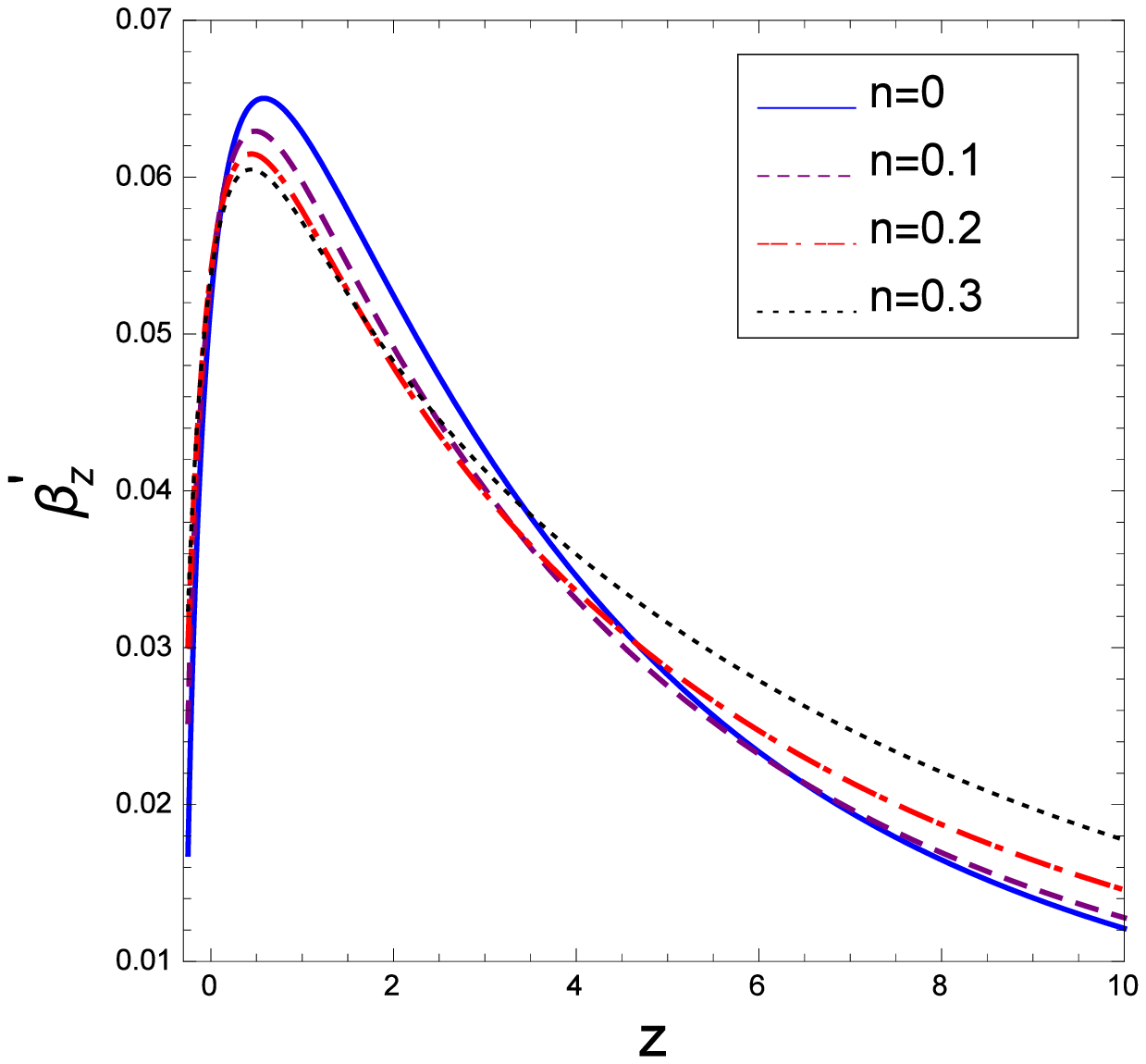}
\caption{The change of the observable acceleration along axis $\beta'_z$ with $z$ for different $n$. The figure is  plotted with LogLog scale in the left panel and with linear scale in the right one. }
\label{Fig4}
\end{figure}

In Fig.(\ref{Fig1}), we plot numerically the changes of $\rho$ with $z$ for different $n$. In order to make a comparison with the case in the Kerr black hole spacetime,  we take the values of the parameters
$a=0.877004$, $E=10^{6}$, and $Q=9.238849$ as in Refs.\cite{Gariel1,Anz2}. It is shown that the curves describing the change of $\rho$ with $z$ have the similar shapes for the different $n$. However, Fig.(\ref{Fig1}) tells us that the values of $\rho$ increases with the NUT charge $n$, which means that the larger $n$ makes the collimated jet's trajectory farther from the $z-$axis. In Figs.(\ref{Fig2})-(\ref{Fig4}), we also present the observable velocity $\beta_{z}(z)$, the proper acceleration $\ddot z(z)$ and the observable acceleration $\beta'_z$ for different $n$. With the increase of the NUT charge $n$,  the observable velocity $\beta_{z}(z)$ increases monotonously. For the proper acceleration $\ddot z(z)$, we find that it decreases with $n$ for the smaller $z$,  while it increases for larger $z$. Thus, in the region near $z-$axis, the effects of the NUT charge $n$ on the proper acceleration of the particle closing to the center of black hole is different from those particles far from black hole. For the Kerr case (i.e., $n=0$), we find that there exist a region in which  the proper acceleration is weakly negative. However, in the KTN black hole spacetime, the region with the negative proper acceleration disappears quickly with the emergence of the NUT charge $n$.
Moreover, the observable acceleration along axis $\beta'_z$ increases with $n $ in the regions with the small $z$ and the large $z$, but it decreases in the intermediate region of $z$. Thus, the NUT charge modifies the repulsive effect for the particles moving along $z-$axis in  the black hole spacetime.

\section{\bf Summary}

In this paper, we have investigated the acceleration of the unbound high energy particles moving along the symmetry axis of the KTN metric in the small $n$ case, and then study the effect of the NUT charge on the acceleration in different regions of the spacetime. We find that the presence of the NUT charge $n$ modifies the condition of the timelike particle leaving the ergosphere at $\rho=0$. Moreover, for the unbound high energy particles leaving the ergosphere at $\rho=0$, the NUT charge enhances the acceleration $\ddot{z}$ along $z$ axis. However, for the particle leaving the ergosphere at $\rho=\rho_e$ or moving near the equatorial plane, the effects of the NUT charge on the acceleration along $z-$axis become more complicated, and the relationship between the repulsive effects and the NUT charge also relies on the Carter constant $Q$ and the velocity along $z$-direction $\dot{z}$ of the unbound high energy particles. We also present numerically the dependence of the observable velocity and acceleration on the NUT charge for the unbound particles in the Kerr-Taub-NUT spacetime.
With the increase of the NUT charge $n$,  the observable velocity $\beta_{z}(z)$ increases monotonously. However, the effects of the NUT charge $n$ on the proper acceleration of the particle in the region near $z-$axis depend on the value of $z$ since the proper acceleration $\ddot z(z)$ decreases with $n$ for the smaller $z$ and increases for the larger $z$.
Moreover, the observable acceleration along axis $\beta'_z$ increases with $n $ in the regions with the small $z$ and the large $z$, but it decreases in the intermediate region of $z$. 
Our result at least in principle show a potential effect arising from gravitomagnetic monopole moment on the particle acceleration in the purely gravitational effect model connected with astrophysical jets.

\section{\bf Acknowledgments}

This work was partially supported by the Scientific Research
Fund of Hunan Provincial Education Department Grant
No. 17A124. J. J.'s work was partially supported by
the National Natural Science Foundation of China under
Grant No. 11475061.

\end{document}